# Closed Timelike Curves, Singularities and Causality: A Survey from Gödel to Chronological Protection

Jean-Pierre Luminet

Aix-Marseille Université, CNRS, Laboratoire d'Astrophysique de Marseille , France;
Centre de Physique Théorique de Marseille (France)
Observatoire de Paris, LUTH (France)
jean-pierre.luminet@lam.fr

**Abstract:** I give a historical survey of the discussions about the existence of closed timelike curves in general relativistic models of the universe, opening the physical possibility of time travel in the past, as first recognized by K. Gödel in his rotating universe model of 1949. I emphasize that journeying into the past is intimately linked to spacetime models devoid of timelike singularities. Since such singularities arise as an inevitable consequence of the equations of general relativity given physically reasonable assumptions, time travel in the past becomes possible only when one or another of these assumptions is violated. It is the case with wormhole-type solutions. S. Hawking and other authors have tried to save the paradoxical consequences of time travel in the past by advocating physical mechanisms of chronological protection; however, such mechanisms remain presently unknown, even when quantum fluctuations near horizons are taken into account. I close the survey by a brief and pedestrian discussion of Causal Dynamical Triangulations, an approach to quantum gravity in which causality plays a seminal role.

**Keywords:** time travel; closed timelike curves; singularities; wormholes; Gödel's universe; chronological protection; causal dynamical triangulations

## 1. Introduction

In 1949, the mathematician and logician Kurt Gödel, who had previously demonstrated the incompleteness theorems that broke ground in logic, mathematics, and philosophy, became interested in the theory of general relativity of Albert Einstein, of which he became a close colleague at the Institute for Advanced Study at Princeton. He then discovered an exact solution of the gravitational field equations describing a rotating universe model [1].

Gödel's universe is infinite, non-expanding, and filled with an idealized, homogenous perfect fluid. It rotates to stay balanced against gravitational collapse, the angular velocity outward juxtaposed to the gravitational pull inward. Gödel explained that matter rotates relative to the compass of inertia with the angular velocity $2(\pi G\rho)^{½}$, where ρ is the mean density of matter, and G, Newton's gravitational constant. However, unlike a spinning top, Gödel's universe does not rotate about a privileged geometrical axis. The local inertial frames rotate with respect to a distant frame defined by faraway galaxies so that every observer sees himself at the center of rotation, hardly an unusual situation. More extraordinarily, in Gödel's rotating universe, there are space-time trajectories that return to their starting point, namely closed timelike curves (CTCs). Such curves remain confined locally to their future light cones, and an object traveling along a CTC never moves faster than the local speed of light, so that CTCs do not all violate the laws of special relativity, while representing possible paths for material objects. This is a very strange result because a traveler could journey into the future but arrive in the past, accompanied by all the paradoxes that arise from a possible violation of the principle of causality.

At the time when Gödel published his article, time travel in the past was already a staple of science fiction, and some of its associated paradoxes had already been discussed in several issues of the 1917 edition of *Electrical Experimenter* (which would become later the infamous *Amazing Stories* magazine). The best known of the paradoxes is the grandfather paradox. As far as I know, it was first formulated in 1936 by Catherine Moore in a short story entitled "Tryst in Time," in which a time traveler tries to kill his own grandfather [2].

The French writer René Barjavel used the same idea in the final scene of his famous novel *The Reckless Traveler*, published in 1944 (for an English translation, see [3]). Eager to change the course of history, the hero travels back in time to the Napoleonic era and inadvertently kills a man who is, in fact, one of his ancestors. Would he survive the encounter if he broke the causal chain leading to his own existence? The 1958 edition also includes a postscript entitled "To Be *and* Not to Be," in which the novelist specifies the nature of the time travel paradox. It seems that Barjavel, an avid reader of popular science literature, was aware of Schrödinger's thought experiment involving a cat at once half-dead and half-alive, as well as the many-worlds interpretation of measurement in quantum mechanics proposed by Hugh Everett in 1957 [4].

Some authors have suggested that Everett's theory offers a possible solution to the grandfather paradox: by killing his ancestor, the traveler would bring about a change in the future, resulting in a bifurcation of space-time along several different causal lines. Such a solution is unsatisfactory because it is based on a misunderstanding of the many-worlds hypothesis. Everett proposed an interpretation of the measurement process in quantum physics, according to which all states in superposition continue to exist after measurement but in disjoint universes—contrary to the standard Copenhagen interpretation—where the measurement sets the system in a unique classical state. But the murder of an ancestor is not a quantum process. Moreover, even assuming a bifurcation of the universe generated by the elimination of some ancestor, the paradox would be logically resolved only if different causal lines interacted in a very particular way, which is also contrary to Everett's theory.

George Gamow had entertained the concept of a universe in global rotation in 1946; he thought it legitimate to generalize to the universe as a whole the observed rotation of local cosmic masses such as planets, stars, and galaxies [5]. Gamow did not derive an exact analytic solution from the equations of general relativity to justify his idea, nor did he mention astronomical observations of the large-scale universe that could support it. The first solutions of Einstein's equations describing the gravitational field generated by a rotating body were proposed by Cornelius Lanczos in 1924 [6] and Willem Jacob van Stockum in 1938 [7]. Van Stockum discussed an infinite rotating dust cylinder, but unlike Gödel's universe later, his model had a geometrically privileged axis. At a critical radius around the cylinder, a light ray emitted orthogonally to the axis of rotation would return to its spatial starting point, but Van Stockum did not realize that further out from the cylinder is a second critical radius, where a light ray emitted orthogonally to the axis returns to its spatio*temporal* starting point, thus forming a CTC. Anyway, although they were not recognized as such by their authors, these models provided the first examples showing a connection between the existence of space-time rotation and time travel.

The works of Lanczos, van Stockum, and Gamow certainly encouraged Gödel to search for an exact relativistic solution describing a rigid rotating universe containing CTC. In addition to the technical feat, which impressed Einstein himself, Gödel sought, above all, to find support for his philosophical point of view on the illusory nature of time. According to Gödel, from the moment when an observer starting from a point in space-time, $p$, travels along a trajectory allowed by the principle of relativity—i.e., confined inside light cones—to his chronological past, $J^-(p)$, time exists relative only to the observer and loses all independent reality. Gödel thought the unreality of time was a physical discovery, and not just a metaphysical option. From his point of view as a logician, Gödel's model of rotating universe, whether realistic or not, was indeed a *possible* universe to the extent that it was an exact mathematical solution of Einstein's field equations. And for Gödel, if time was an illusion in a possible universe, this had also to be true in the real universe.

The philosophical arguments developed by Gödel to deny the physical nature of time have been analyzed by many authors [8]. A thorough study was made by Cassou-Noguès, who had access to thousands of pages and unpublished notes that are preserved at the Institute for Advanced Study of Princeton [9]. In Gödel's papers, Cassou-Noguès saw evidence of a mental system that, while logically consistent, also represented the mind of a scientific genius molded by a bizarre psychology, mixing irrational phobias, superstitions, hypochondria, and paranoia.

The Gödel's rotating universe is not expanding, which is in blatant contradiction to observations of the redshifts of distant galaxies and the cosmic background radiation. More and more drastic experimental constraints have been placed on the possible large-scale vorticity of the universe, starting with Stephen Hawking [10], such that the possibility of a cosmological model in global rotation, although it has been carefully studied, has been almost universally rejected [11].

It is fascinating to note that the philosophical option advocated by Gödel, namely the illusory nature of time, features prominently on the agenda of contemporary physics, with some approaches attempting to reconcile general relativity and quantum field theory. The fundamental equations of loop quantum gravity do not contain a time variable. The theory describes processes in which changes take place not under the action of an identifiable temporal variable but as the result of a non-commutative sequence of spatial operators whose ordered classification simulates an irreversible flow of time (for a topical review, see [12]). As in Gödel's view, but in a very different physico-mathematical context, the experience of passing time would then be relative to the particular conditions in which the observer finds himself. It would also depend on the structure of his cognitive system. Our own perception of time is certainly different from that of other animals. We should not look for a manifestation of our own experience of time in the realm of fundamental physics. On this view, time appears as an emergent quality, underpinned by strata mixing spatial transformations and cognitive processes [13].

## 2. General Relativistic Universe Models with Closed Timelike Curves

In the context of the metrics of space-time solutions to the equations of general relativity, the possibility of travel to the past is related to the existence of CTCs. Gödel's model provides a particular example, but it is far from being unique. The causal structure of space-time is described by the field of light cones at each point. In the Minkowski spacetime of special relativity, free from gravitation, all light cones passing through all events are parallel to one another, somewhat akin to a wheat field in which all the ears are vertical; a CTC would require a departure from the light cones, which is impossible, unless we arbitrarily wrap up the time direction to create trivial CTCs. In general relativity, the situation is much more interesting. By virtue of the equivalence principle, which stipulates the influence of gravitation on all forms of energy, light cones incline and deform according to local curvature. The identification between points of space allowed by a multiply connected topology also deforms the field of light cones to generate topological CTCs [14]. More generally, the existence of CTCs is of a combinated geometrical and topological nature. The reason is that the gravity may constrain the topological possibilities [15]. For instance, Gödel gravity restricts the whole class of possible topologies to the smaller family of not time-orientable spacetimes. In any case, we can conceive of a space-time with strong curvature, or non-trivial global topology, in which the light cones are sufficiently tilted along particular time trajectories to allow them to close in a loop.

In Gödel's rotating universe, as one moves away from a local axis of symmetry the speed of rotation increases, and the field of light cones reaches a point allowing for the formation of timelike loops. Van Stockum's cylinder and its generalization, which was discovered in 1974 by Frank Tipler, allow for the existence of CTCs in the same manner [16]. In fact, Gödel's universe is a particular case of a hyperbolic family of Gödel-type spacetimes [17,18]. These spacetimes are characterized by two physically relevant parameters: $\Omega$ (rotation) and another essential parameter $m$, which is used to divide Gödel spacetimes into 3 disjoint classes: $m^2 > 0$ defines the hyperbolic family which contains Gödel's solution as a particular case, for which $m^2 = 2\,\Omega^2$; the linear class with $m = 0$, which contains only the Som-Raychaudhuri universe, whose CTCs are discussed in [19]; and the trigonometric family, $m^2 < 0$, which contains solutions with alternating causal non-causal regions [20]. For a recent summary, see [21]. CTCs also appear in a family of physically more realistic —e.g., inhomogeneous—Gödel-type universes [22].

Other exact solutions containing closed time curves are the Taub-NUT empty space (for a mathematical description of this vacuum solution, see [23]), the Misner space obtained by introducing a multiconnected torus-like topology into Minkowski space-time [24], or the Gott space representing two parallel cosmic strings of infinite length that intersect at relativistic speed [25]. But the most famous solution is by far Kerr's space-time describing a rotating black hole (for a comprehensive mathematical description, see [26]).

These space-times include chronal regions devoid of CTCs, and one or more achronal regions that contain CTCs; their common boundary is called the chronological horizon [27]. The Kerr solution has two event horizons denoted $r_+$ and $r_-$ respectively. The regions $r > r_+$ and $r_+ > r > r_-$ are chronal, while the region $r < r_-$ is achronal; the inner event horizon $r_-$ plays the role of chronological boundary. A timelike curve from a point $p$ of the chronal region outside the black hole can successively cross the two event horizons $r_+$ and $r_-$, cross the achronal region $r < r_-$, and emerge from the black hole to arrive in the chronological past $J^-(p)$, thus achieving a CTC of the type referred to as a wormhole.

The first wormhole solution was published in 1916 by Ludwig Flamm [28,29]. It was studied in 1935 by Einstein and Nathan Rosen as part of Schwarzschild's solution describing a static spherical black hole and was referred to as an Einstein-Rosen Bridge [30]. The term "wormhole" first appeared in a famous article published in 1957 by Charles Misner and John Wheeler [31]. The introduction of the Kruskal coordinate system in the 1960s made it fully possible to describe the geometrical and topological structure of Schwarzschild's space-time and to show that the existence of a timelike singularity at $r = 0$ in the middle of Einstein-Rosen Bridge renders a wormhole untraversable.

The question then arose whether there were more general solutions that allowed for CTCs in association with traversable wormholes. Answers in the affirmative could be found from Kerr's solutions for a rotating black hole, the Reissner-Nordström metric for an electrically charged black hole, and the so-called Lorentzian wormholes lined with exotic material, that is to say, negative energy density [32]. If these space-times have many singularities, they are not timelike but spacelike. Kerr's black holes have a ring-shaped singularity lying in the equatorial plane, and perfectly lawful particle paths can either fly over it or pass through it without touching it. The models of Gödel, van Stockum, Tipler, Misner, and Gott are devoid of any singularity. The existence of CTCs is, in fact, contingent on the absence of timelike singularities.

## 3. Singularities and Time Travel

The crucial question of the occurrence of singularities in (classical) general relativity was dealt with for the first time in a 1933 article by Georges Lemaître, which unfortunately went mostly unnoticed at the time because it was published in French [33]. Next, in 1955, Raychaudhuri provided the equation (now named after him) describing geodesic focusing and its possible consequences as physical singularities [34]. Ten years later, Roger Penrose [35] gave the first modern proof of the occurrence of a future singularity in gravitational collapse, introducing the fundamental notion of closed trapped surface (as it was recently recognized by the Nobel prize in physics 2020), while Stephen Hawking studied its analogue by reversing time, namely a singularity of the past reached in big bang cosmological solutions. The classic reference is [36]; see also the more recent [37], which contains technical details and important historical notes. Using differential topology, these so-called "singularity theorems" proved that the appearance of timelike singularities was generically inevitable, given some simple physical hypotheses. These are grouped into two major classes. The first one places certain conditions on space-time: the existence of a Cauchy surface on which it is possible to define initial conditions whose development unequivocally determine the past and the future, or a limit on the rotation velocity of the universe or the bodies it contains. The second consists of various constraints on the positivity of energy—so-called strong, weak, or dominant conditions.

These important singularity theorems are similar to theorems governing rays in geometrical optics. A singularity is similar to a point of convergence in an optical system. The event horizon of a black hole is like a convergent lens: all the particles and light trajectories that cross it are focused on a single point, which corresponds to a central singularity. In the same way, the appearance of a singularity in big bang cosmology is established from the convergence of all timelike trajectories as one goes back in the past.

By marking an interruption of time for matter and light trajectories, singularities prevent these trajectories from closing in a loop and violating causality. In classical general relativity, i.e., without quantum corrections, the possibility of traveling in the past is therefore necessarily related to a violation of one or another of the hypotheses concerning the occurrence of timelike singularities.

According to general relativity, a wide variety of solutions give rise to CTCs [38]. But in each case, one or another of the hypotheses governing the occurrence of singularities is not satisfied. The cylinders of infinite length in relativistic rotation are not physically relevant, our universe does not rotate as fast as the Gödel solution requires, cosmic strings of infinite length are unrealistic, traversable wormholes violate the positivity energy condition, and so on. The debate might simply end with a decree that nature forbids the formation of CTCs. Nonetheless, one can well imagine a technologically advanced civilization that is able to artificially create closed timelike curves in order to travel in time. This is the scenario depicted in Carl Sagan's novel *Contact*. An ancient extraterrestrial race transmits a radio message that contains the plans for a time machine. When travelling from star to star, the machine uses a space-time shortcut in the form of a wormhole, an approach that violates the singularity theorems. Sagan asked his colleague Kip Thorne to find a workaround. Thorne thus embarked upon a parallel career as consultant and scientific advisor for novelists and filmmakers (in 2014, Thorne was the scientific advisor for the film *Interstellar*, see [39]). He replied to Sagan that it was sufficient to violate the positive energy conditions guaranteeing the appearance of singularities by introducing a form of exotic energy with negative pressure. All the usual forms of matter obey positive energy conditions. No body made of classical material in gravitational implosion, such as a black hole, or even the big bang, can generate an achronal region. But quantum field theory offers a means of escape. Quantum vacuum energy fluctuations are described as bubbling pairs of particles and antiparticles of all types and masses (electrons, quarks, photons, mini black holes, and so on), constantly appearing and disappearing. These are felt, in particular, as part of the effect predicted by Hendrik Casimir in 1948 and verified experimentally by Steve Lamoreaux in 1997. The introduction into the quantum vacuum of two weakly separated conductive plates modifies the energy value of a vacuum, which then behaves as if it possesses a negative energy.

In 1988, Thorne and his collaborators suggested that not only was it theoretically possible to open a passage between two points of space and to travel almost instantaneously between the stars but that we might also be able to travel in time [40]. Having opened and stabilized a traversable wormhole, one can imagine leaving one of its mouths on earth while the second is carried inside a spaceship in relativistic flight, making a round trip of, say, ten thousand years for an observer remaining behind on earth. If the journey is undertaken at speeds close to those of light, the passage of time onboard will be only a few days or weeks. A time lag will then arise between the two mouths of the wormhole. This would allow for time travel of up to ten thousand years back into the past for those onboard the ship or up to ten thousand years into the future for those left behind on earth.

In the same order of ideas involving exotic matter with negative energy density, the Alcubierre warp drive [41] and the Krasnikov tube [42] are solutions of classical Einstein equations leading to theoretical time machines.

**4. How to Escape CTCs?**

Geometries of this type offer fascinating possibilities, but they violate causality, a basic physical law which is believed to be fundamental. Proposals that allow for backwards time travel but prevent time paradoxes were first suggested by Novikov [43], then developed by collaborators [44]. Their so-called "self-consistency principle" asserted that if travel along a CTC exists and would cause a paradox or any change to the past whatsoever, then the probability of such travel is zero. In other words, it demands that local timeline curves are admissible only when they are globally "self-consistent", i.e., if geometry is consistent with a topology that does not allow for CTCs. Thus, this principle appears as an *ad hoc* global topological constraint on admissible local solutions, which ultimately forbids time-non-orientable spacetime manifolds. It just adds a topological restriction with no genuine physical motivation.

The orientability of time is not a physical law but an open question in principle falsifiable by experiments in order to gain physical consistency. Indications could come from quantum effects, which were not really taken into account in the precedent discussions. In quantum field theory without gravity, causality is closely related to the principle of locality, namely that an object is directly influenced only by its immediate surroundings, thus forbidding the possibility of instantaneous "action at a distance".

However, the locality principle is disputed according to various interpretations of quantum mechanics, including when discussing quantum entanglement showing a violation of Bell's inequalities [45]. Lacking a full integration of general relativity and quantum mechanics in a theory of quantum gravity, semi-classical gravity provides an approximate method for modeling quantum fields in the curved spacetime of general relativity. Quantum systems traversing CTCs have been studied in this context [46], and experimental simulation of CTCs have been undertaken [47].

Using such a semi-classical approach, Hawking tried to counter the eventuality of a traversable wormhole by proposing, in 1992, a "chronological protection conjecture" [48]. There should exist, according to the conjecture, a physical mechanism capable of preventing CTCs from forming in *any conceivable* circumstances, whether natural or artificial. Hawking argued that the achronal regions of the Misner, Taub-NUT, and Kerr spaces are classically unstable; particles and fields falling into a Kerr black hole, or traveling at relativistic speed in Taub-NUT and Misner spaces, see their spectral shift diverge toward blue as the chronological horizon approaches. Thus, it seemed to him reasonable to think that the associated energy density, which is also divergent, had sufficient feedback over space-time to prevent the formation of CTCs.

One possible method for finding a universal protection mechanism can indeed be found in the quantum instability of time horizons. This can be described as a stacking of the quantum fluctuations of the vacuum in the vicinity of the chronological horizon, so that the fluctuations have a non-zero renormalized energy density that diverges as the horizon approaches. In turn, semi-classical Einstein equations suggest that this energy should distort the space-time geometry in order to protect the timeline. Such a feedback mechanism is the quantum analogue of the well-known Larsen effect in acoustics.

In a first study, when the calculation for a beam of radiation entering a wormhole mouth was done taking account of vacuum fluctuations, it was found that the beam would spontaneously refocus before reaching the other mouth, suggesting that the "pileup effect" becomes large enough to destroy the wormhole [49]. Uncertainty about this conclusion, however, remained because the semi-classical calculations indicate that the pileup would only drive the energy density to infinity for an infinitesimal moment of time, after which the energy density would die down [50]. But semi-classical gravity is considered unreliable for large energy densities or short time periods near the Planck scale, and a complete theory of quantum gravity is needed for accurate predictions. So, it remained uncertain whether quantum-gravitational effects might prevent the energy density from growing large enough to destroy the wormhole.

Indeed, subsequent works in semi-classical gravity provided examples of spacetimes with CTCs where the energy density due to vacuum fluctuations does not approach infinity in the region of spacetime outside the Cauchy horizon [51]. But in 1997, a general proof was given demonstrating that according to semi-classical gravity, the energy of the quantum field (more precisely, the expectation value of the quantum stress-energy tensor) must always be either infinite or undefined on the horizon itself [52]. Both cases indicate that semi-classical methods become unreliable at the horizon because quantum gravity effects would be important there; this is consistent with the possibility that such effects would always prevent time machines from forming. Nevertheless, the chronological protection conjecture remains unproven as no rigorous proof could be formulated in all the spaces that could conceivably host CTCs. It may be the case that chronology is not *always* protected at macroscopic scales, and even if it were, quantum gravity could give rise to nonzero probability amplitudes allowing CTC formation on a microscopic scale–for the latest developments, see [53,54]. Whether or not the chronology protection principle holds, Hawking and Penrose have pointed out that too severe causality assumptions, such as global hyperbolicity, could risk "ruling out something that gravity is trying to tell us" [55].

A definite theoretical decision on the status of the chronology protection conjecture would require a full theory of quantum gravity as opposed to semi-classical methods. So, what do presently available theories of quantum gravity say? Although none of the various approaches for quantizing gravity has proven to be fully consistent, the general tendency is that quantum effects should prevent the occurrence of timelike singularities, thus removing one of the restrictions to the physical existence of CTCs. In the general framework of string theory, geometries with CTCs have resurfaced as solutions to its low energy equations of motion and time travel appeared to be possible in these geometries. However, the situation is far from being clear as it was suggested that stringy effects should prohibit their construction. A (completely unrealistic) example is provided by the extremal supersymmetric five-dimensional charged spinning "BMPV" black hole [56], which contains causality violating regions. However, taking account of stringy effects, it was shown how the geometry in these achronal regions had to be corrected and that, once corrected, causality is preserved. More precisely, tracking the chronology protection conjecture in the dual conformal field theory reveals that the absence of CTCs in the geometry coincides with the preservation of unitarity in the conformal field theory [57]. There are more arguments that seem to support chronology protection in string theory [58,59]. Moreover, Gödel's universe seems to lose its closed timelike curves when modeled in string theory [60].

But string theory is not a complete theory of quantum gravity. Experimental observation of closed timelike curves would of course demonstrate the chronology protection conjecture to be false, but, short of that, if physicists had a theory of quantum gravity whose predictions had been well-confirmed in other areas, this would give them a significant degree of confidence about the possibility of time travel or not. None of the approaches to quantum gravity can do so. But the preservation of causality serves as a basic element of construction for two of them, namely the Causal Sets theory [61] and the Causal Dynamical Triangulations (CDT) theory [62]. I will end this review by commenting on CDT in a pedestrian approach [63].

## 5. Focus on CDT

According to John Wheeler's quantum geometrodynamics, near the Planck scale the structure of space-time would be analogous to a turbulent "foam", constantly changing due to quantum and topological fluctuations. One of the major difficulties of the approach is to describe how the microscopic foam of space-time can evolve and materialize on a macroscopic scale into a four-dimensional universe similar to the one we live in today. CDT theory attempts to do this by adapting Regge's computation to quantum space-time. Its very name comes from the fact that the theory uses a dynamically varying process of triangulation of space that follows deterministic rules respecting causality.

Previous attempts to explain the quantum structure of space-time had very limited success. They were based on Euclidean quantum gravity and a fundamental principle of quantum mechanics: superposition. Applying the principle of superposition to the entire universe requires considering the very vast set of all possible geometries for space-time, these geometries playing the role of paths. Calculating path integrals from this set is extremely difficult, if only to simply identify all paths, i.e., all possible geometries.

Euclidean quantum gravity made a great technical leap forward during the 1980–1990s with the development of powerful computer simulations. Initially, this involved replacing continuous geometries with discontinuous approximations. Simulations used simplexes as four-dimensional building blocks. Unfortunately, they produced either "monster" universes with an infinite number of dimensions or, on the contrary, "minimalist" universes with barely two dimensions. They were clearly missing an important ingredient.

The essential improvement that dynamical triangulations brought to Euclidean quantum gravity was to incorporate a causal structure of space-time from the outset. The latter, of dimension 4, is considered as a succession of spatial slices indexed by a discretized temporal variable, t. Each spatial slice is decomposed into regular simplexes of dimension 3, and the connection between the slices is made by a piecewise linear variety of 4-simplexes. Instead of a smooth variety, a network of triangulation nodes is obtained, where the space is locally flat within each simplex but globally curved. The flat faces that define each simplex can represent either a spatial or temporal extent, depending on how they reside in a given time slice, or connect a vertex at time *t* to another vertex at time *t* + 1.

Compared to previous attempts at triangulation, the crucial development of CDT theory [64] was to add the constraint that the network of simplexes must evolve in such a way as to preserve causality. Each simplex is first assigned a time arrow from the past to the future. Then the rules of causal gluing are applied: two simplexes can only be glued together if their arrows point in the same direction. The simplexes must share the same notion of a time that unfolds constantly in the direction of the arrows and never stands still or goes backwards. Space retains its general shape as time progresses; it cannot divide into disconnected pieces or create CCTs and wormholes.

Computer calculations of the evolution of a large causal superposition of 4-simplexes resulted in a large-scale space-time of dimension 4.02 ± 0.1. [65]. In short, if we consider empty space-time as an immaterial substance composed of a very large number of tiny building blocks and then let these microscopic blocks interact with each other according to the simple rules dictated by gravity and quantum theory, they spontaneously organize themselves into a collective whole that resembles the universe observed today.

The next step in CDT theory was to study the shape of space-time on a very large scale and to check its agreement with reality, i.e., the cosmological solutions of general relativity. This test is very difficult to pass in non-perturbative models of quantum gravity, which do not assume a particular default shape for space-time. In fact, it is so difficult that most approaches to quantum gravity—including string theory—are not sufficiently advanced to pass it. CDT theory can do it provided that simulations include the cosmological constant from the outset. In this case, the emerging space-time has De Sitter's geometry, which is the exact solution of Einstein's equations for a universe containing only the cosmological constant.

It is truly remarkable that by assembling microscopic building blocks in an essentially random manner, without taking into account any symmetry or privileged geometric structure, the simulations result in a space-time that has the very symmetrical structure of De Sitter's universe. Moreover, this classical De Sitter geometry remains valid up to a scale of about $10^{-33}$ cm, although quantum fluctuations become increasingly important at this level.

The fact that one can trust De Sitter's classical approximation at such short distances is rather astonishing. This has important implications for the universe, both very early in its history and far into the future. At both extremes, the universe is indeed dominated by the quantum vacuum. In the beginning, gravitational quantum fluctuations were so huge that matter played no role. In a few billion years' time, if the accelerated expansion of the universe continues at an exponential rate (depending on the equation of state of dark energy), matter will be so diluted that it will again play no role. The CDT theory can explain the structure of the universe in these two extreme regimes.

Once this classical test had been successfully passed, CDT researchers moved on to another type of calculation to probe the truly quantum structure of space-time, which Einstein's classical theory and De Sitter's solution fail to capture. To do this, a revealing numerical simulation consists of reproducing a diffusion process. In simple terms, it is a matter of letting the analogue of a drop of ink "fall" into the superimposed universe and observing how it propagates and scatters under the effect of quantum fluctuations. By measuring the size of the ink cloud after some amount of time, it is possible to determine the number of dimensions in space. The result was surprising: this number depends on the scale. If diffusion is observed over a short period of time, space-time seems to have a different number of dimensions than if diffusion is allowed to act over a longer period of time. It is difficult to imagine how space-time can thus "smoothly" change dimension according to the resolution of the microscope. It would seem that a very small object "experiments" with space-time in a profoundly different way than a large object. For the microscopic object, the universe would look like a fractal.

For example, numerical simulations on scattering have shown that at Planck's length, quantum fluctuations in space-time become so strong that the classical and intuitive notions of geometry completely collapse. The number of dimensions decreases from the classical value of four to a value of about two [66]. The fact that it is not exactly equal to two denotes the fractal nature of space. Nevertheless, space-time remains continuous and has no more microscopic wormholes than macroscopic ones. Chronological protection is thus ensured at all scales.

Like any candidate theory for quantum gravitation, the Holy Grail is to predict observable consequences derived from the microscopic quantum structure. This is the ultimate criterion for deciding whether one has a correct theory of quantum gravity. The future will tell whether the CDT theory succeeds in doing so.

To conclude, from our current knowledge of physics, journeys to the past seem reasonably impossible. Time machines seem doomed to remain forever in the realm of cinema and science fiction.

Still, we must be careful not to prejudge the surprises that the physics of tomorrow might offer.

**Acknowledgments:** The author is very much obliged with Lorenzo Iorio for his invitation to submit an invited contribution to the 5-year commemorative Special Issue of the journal *Universe*.